\documentstyle[prb,preprint,aps]{revtex}
\begin{document}
\draft

\title{Vibrational Study of $^{13}$C-enriched C$_{60}$ Crystals}

\author{Michael C. Martin$^{1}$, J. Fabian$^{1}$, J. Godard$^{2}$,
P. Bernier$^{3}$, J.M. Lambert$^3$, and L. Mihaly$^{1}$}
\address{$^{1}$Department of Physics, SUNY at Stony Brook, Stony Brook, NY
11794-3800\\
$^{2}$Laboratoire de Physique des Solides, Universit\'e Paris-Sud,
F-91405 Orsay, France\\
$^{3}$Groupe de Dynamique des Phases Condens\'ees, Universit\'e des
Sciences et Techniques du Languedoc, 9F-34060 Montpellier, France}

\date{14 September 1994, revised 1 November 1994, Accepted to Phys. Rev. B}
\maketitle

\begin{abstract}
The infrared (IR) spectrum of solid C$_{60}$ exhibits many weak vibrational
modes.  Symmetry breaking due to $^{13}$C isotopes provides a possible route
for optically activating IR-silent vibrational modes.  Experimental
spectra and a semi-empirical theory on natural abundance and $^{13}$C-enriched
single crystals of C$_{60}$ are presented.  By comparing the
experimental results with the theoretical results, we exclude this
isotopic activation mechanism from the explanation for weakly active
fundamentals in the spectra.
\end{abstract}
\pacs{PACS: 74.30.-j, 74.30.Jw, 74.70.Wz}


The availability of high quality C$_{60}$ single crystals has provided an
opportunity for the detailed study of weak vibrational features in
the infrared (IR) and Raman spectra of this molecular solid.  Recent
studies identified some of these modes as being weakly activated
symmetry-forbidden modes in IR \cite{eklundir,c60modes,kamaras} and
Raman \cite{eklundraman,vanloos} measurements.  Second-order combination
modes were also observed in these spectra.  These results enabled the
authors to use group theoretical arguments to extract candidates for
all 46 vibrational modes of C$_{60}$ from the experimental
data.\cite{c60modes,eklundraman}

Since the icosahedral ($I_h$) symmetry of the isolated fullerene molecule
allows for only four IR active modes, this symmetry must be broken by some
agent.  Possible mechanisms for activating such ``silent" modes include
$^{13}$C isotopic impurities, Van der Waals forces from other C$_{60}$
molecules (crystal field effects), anharmonicity, electric field gradients
at surface boundaries, impurities, and dislocations.  Several previous works
\cite{eklundir,c60modes,eklundraman,eklundisotope} have argued that
$^{13}$C isotope substitution is a likely agent for activating silent modes
since for the natural abundance of $^{13}$C, about one half of all C$_{60}$
molecules would have at least one $^{13}$C which would break the $I_h$
symmetry and activate all the vibrational modes.  In the present study,
we experimentally and theoretically explore this idea by studying C$_{60}$
crystals made from natural abundance and $^{13}$C-enriched C$_{60}$.

The natural abundance $^{13}$C single crystals were sublimated as described
previously \cite{c60modes} from well cleaned C$_{60}$ powder commercially
produced.\cite{ses}  Typical crystals have dimensions $\sim 1\times 1
\times 0.5$mm.  The 8\% $^{13}$C-enriched crystals were sublimated from
enriched soot prepared as reported earlier \cite{8prep} and their
chemical and crystallographic quality were checked by high-resolution
$^{13}$C solid state nuclear magnetic resonance (NMR).\cite{8nmr}  The
isotope enrichment was determined by mass spectrometry, showing that
the mass distribution follows a Poisson function.
A Bomem MB-155 FTIR spectrometer was used to obtain 2cm$^{-1}$ resolution
IR spectra in the 400-5000cm$^{-1}$ energy range.  The infrared spectra
presented here were obtained at room temperature with the crystals
mounted over a small aperture with a minimal amount of silver paste.  The
spectra are referenced to spectra obtained through the empty aperture.
The IR spectra of such thick crystals has been shown to be independent
of atmosphere\cite{c60modes}, so for experimental
simplicity the crystals in this study were exposed to air.

The probability that a C$_{60}$ molecule has $n$ $^{13}$C atoms is given by
the Poisson distribution\cite{eklundisotope}, $p_n$, in agreement with the
mass spectrometry of the 8\%-enriched $^{13}$C samples.\cite{8prep}  This
function is presented in Figure \ref{dist} for C$_{60}$ molecules made from
natural abundance carbon (1.108\% $^{13}$C) and from 8\% $^{13}$C-enriched
carbon.  The natural abundance probability curve shows that about one third
of the molecules will have exactly one $^{13}$C atom which will weakly
break the icosohedral symmetry of pure $^{12}$C$_{60}$.  From the 8\%
probability function we learn that only about 1\% of C$_{60}$ molecules
will be pure $^{12}$C$_{60}$; most will have 3 to 7 $^{13}$C atoms per
fullerene.

The IR data are presented in Figure \ref{spectra}.  The uppermost curve is
the transmission of a natural abundance C$_{60}$ single crystal, as published
previously.\cite{c60modes}  The four IR-active $F_{1u}$ vibrational modes
are observed to be completely saturated around their known positions of
527, 576, 1182, and 1427cm$^{-1}$.  All other observed
absorptions are due to IR-forbidden fundamental vibrations or second (or
higher) order modes \cite{c60modes,eklundir,eklundraman,kamaras} which
have become weakly active in the crystal.  One of the weak modes at
710cm$^{-1}$ was fitted in a previous study \cite{c60modes}; its intensity
(proportional to the square of the plasma frequency of the Lorentz
oscillator, $\omega _p^2$) is approximately one order of magnitude
smaller than that of a typical $F_{1u}$ mode.

The middle curve of Figure \ref{spectra} displays the IR spectrum of an
8\% $^{13}$C-enriched C$_{60}$ crystal.  When comparing this spectrum with
that of the natural abundance crystal (top curve of Fig. \ref{spectra}),
the primary changes observed are that all of the visible modes have
broadened and slightly softened.  Note, however, that the relative
intensities of the weakly active fundamental and higher order vibrational
modes do not change significantly.  Nearly all the features of
the 8\% enriched sample transmission spectrum can be modeled by simply
broadening the spectrum obtained for the natural abundance crystal using
the functions from Figure \ref{dist} and assuming the energy of a mode
is proportional to $1/\sqrt{m}$ (like in a simple ``ball and spring" model).
(This relation between mode broadening and mass distribution was described
in greater detail for the $A_g(2)$ Raman fundamental by Guha
{\it et al.}.\cite{guha})  Carrying this procedure out by applying the
8\% Poisson distribution point by point to the top spectra of Fig.
\ref{spectra}, we obtain the bottom curve of Figure \ref{spectra}.
A careful comparison of the measured 8\% enriched spectrum and the
spectrum generated from the broadening reveals that
they are indeed very well matched, indicating that $^{13}$C substitutions
change the frequency of the mode, but do not influence the intensity.
Raman spectra obtained on these samples in which forbidden and higher order
modes are observed give the same results.\cite{armand}

If the fundamental modes are being activated by symmetry breaking due to
isotopic substitutions, one would expect the $^{13}$C-enriched samples to
have greater symmetry breaking, and therefore greater IR-activity for
originally silent modes.  To test this, we used a theoretical model of
C$_{60}$ based on the bond-charge model of Onida and Benedek \cite{model}
(the only change was to use the better experimental values \cite{bond}
of 1.4\AA \ for the hexagon-hexagon (6-6) bond length and 1.45\AA \ for
the pentagon-hexagon (5-6) bond length).  The
vibrational frequencies calculated with this model are within 3\% of the
experimental IR and Raman mode frequencies.

The calculation of the IR intensity of a given mode requires, in
principle, the knowledge of the electronic eigenstates of the deformed
molecule.   To avoid the numerical complexity of such an approach we used an
approximate procedure, based on the experimentally known IR intensities of
the four allowed $F_{1u}$ modes and on the calculated vibrational
eigenvectors of the isotopically changed molecule.  When expanded in terms
of the original eigenvectors of the pure $^{12}$C molecule, some or all
of the new eigenvectors had components in the directions of the original
$F_{1u}$ eigenvectors.  Each degenerate species of the unperturbed $F_{1u}$
eigenvectors are chosen to carry their dipole moments in the Cartesian
$x,y$, and $z$ directions.  The intensity of each of the four IR-active
$F_{1u}$ modes was then taken from experiment \cite{c60dope} and are denoted
$A_1, A_2, A_3$, and $A_4$.  The IR-intensity of mode $l$ is then calculated
using
\begin{equation}
I_l=\sum_{i=1}^3 \left( \sum_{j=1}^4 \sqrt{A_j}\ \tilde{Q}_{jl}^i
\right) ^2 \ ,
\label{eqn1}
\end{equation}
where $\tilde{Q}_{jl}^i$ is the component of eigenmode $l$ in the direction
of the the $i^{\rm th}$ degenerate mode of the original $F_{1u}(j)$
eigenvector; $i$ sums over the triple degeneracies and $j$ sums over the
four $F_{1u}$ modes.  The isotopic content is modeled by a Monte Carlo
simulation: each of the 60 carbon atoms is given a $^{13}$C or a $^{12}$C
mass with appropriate probabilities representing the natural abundance or
$^{13}$C-enrichment (1.108\% and 8\% $^{13}$C, respectively).  The dynamical
matrix for the resultant molecule was then diagonalized to calculate the
eigenvectors, and Eqn. \ref{eqn1} was applied.  For a given isotope
concentration, this process is repeated until a general sampling
of all possible $^{13}$C configurations is calculated and the results are
averaged together.   For the natural abundance case, there are relatively
few possible configurations, but for the 8\% $^{13}$C-enriched case, there
are a huge number of configurations.  We sampled over 15,000 possible
configurations for the 8\% case, which should suffice to obtain the correct
trends.

We plot the resultant calculated IR-intensity for 1.108\% and 8\% $^{13}$C
C$_{60}$ molecules in Figure \ref{theory}.  The Lorentzian widths of all
modes have been taken to be uniformly 3cm$^{-1}$.  The four $F_{1u}$ modes
in Figure \ref{theory} have peak heights of approximately 2000
($\omega _p^2\sim 5000$cm$^{-1}$), so the intensities of the weakly activated
modes are approximately a factor of $10^4$ smaller.  This is in contrast to
the experimentally observed intensity ratio of only a factor of $\sim 10$.
Furthermore, the average intensity of the weakly activated modes increases
by a factor of 4.45 in the 8\% enriched calculation compared to the natural
abundance calculation, but no similar increase was seen in the experiment.
Very similar theoretical results are given in a recent detailed calculation
\cite{tanaka} of the effects of isotopes on the vibrations of C$_{60}$.

We must therefore conclude that isotopic activation of previously
silent modes is not the symmetry-breaking mechanism in C$_{60}$.

Since the isotope effect is eliminated, the forbidden modes must be
activated by another symmetry breaking mechanism.  Surface effects can
be ruled out since the phenomenon is independent of sample thickness.
Defects in the bulk, like dislocations or vacancies are also less likely,
since the measurements on samples under pressure, or samples subjected to
plastic deformation by pressure do not exhibit a dramatic change in these
weak lines.\cite{c60modes}  Oxygen bound to the fullerenes has been
excluded since the number and strength of the weak vibrational modes are
unchanged when obtained from C$_{60}$ crystals that have never been exposed
to oxygen.\cite{c60modes}  Impurities in the crystals are a possibility,
however we point out that crystals made from different C$_{60}$ sources and
using various growing conditions result in exactly identical IR spectra in
every feature.\cite{eklundir,c60modes,kamaras}  A remaining serious
candidate is the symmetry breaking due to the first neighbor
interactions between the C$_{60}$ molecules in the solid.

The first question is, how is it possible that most of the weak fundamentals
are visible at room temperature, when the X-ray structure is face centered
cubic, with one fullerene molecule per unit cell.\cite{flem91}
In this state the molecules are rotating rapidly and the symmetry breaking
crystal fields fluctuate with this rotation.  To clarify this issue,
we performed a simple model calculation with a fluctuating external
potential breaking the symmetry of an otherwise symmetric arrangement of
atoms.\cite{mihaly}  All other interactions with the surroundings were
represented by a damping constant.  We found that a forbidden mode is
activated ({\it i.e.} the system absorbs energy from the external
oscillating electric field in a resonance like fashion) as long as the
correlation time of the fluctuations is not shorter than the lifetime of
the resonance (which, in our calculation, was determined by the damping
constant).

To apply this idea to the present case, we need to know the natural
lifetime of a particular vibrational mode, and the correlation time of the
fluctuations of neighbors.  The first parameter can be estimated from
the linewidth of the well resolved $F_{1u}$ vibrational modes as measured
in detail by Homes {\it et al.}.\cite{homes}  To avoid complications due
to the ``inhomogeneous broadening" we take a linewidth from the measurement
below the orientational ordering transition temperature
($\gamma =0.5$cm$^{-1}$) and
obtain a lifetime of 60psec.  At higher temperatures the lifetime decreases,
therefore this is an upper limit.  On the other hand the rotational
diffusion constant was determined by neutron scattering measurements as
$D_R=1.4\times 10^{10}$sec$^{-1}$ at T=260K, right above the rotational
phase transition.\cite{DR}  This implies that a 60 degree rotation takes
about 130psec, which is indeed longer than the upper limit to our
estimated resonance lifetime.

According the the neutron scattering data, the rotational
diffusion constant increases by a factor of two, to
$2.8\times 10^{10}$sec$^{-1}$ at 520K.  However, the lifetime of the
resonance is also expected to decrease.  In an insulator, the primary channel
of phonon decay is due to the interaction between the phonons, and the
enhanced population of the phonon states leads to shorter lifetimes at
higher temperatures.  These considerations lead to an experimentally
verifiable prediction:  at higher temperatures the ``forbidden" lines
should either disappear (if the lifetime of the resonance has a
weak temperature dependence) or the integrated intensities will
not change while the lines broaden (so that the lifetime stays shorter
than the fluctuation time).  To test this prediction, we measured the
IR spectrum of a crystal at elevated temperatures.  All observed modes
do broaden without a significant change in line intensities up to a
temperature of 650K.

Below T$\sim 250$K, the orientational ordering transition temperature,
the halting of free rotation of the C$_{60}$ molecules leads to a crystal
structure with four fullerenes in the unit cell.\cite{heiney}  In this
structure all the {\it ungerade} modes of the molecule are allowed in the
IR by symmetry \cite{dressel}.  The ideal
Pa$\bar{3}$ structure retains a center of inversion symmetry at the center
of each C$_{60}$ molecule and thus the {\it gerade} modes are still
IR-forbidden.  However there is a ``defect orientation" in
the actual structure which will break this inversion symmetry in a
significant portion of the C$_{60}$ molecules.\cite{david}

According to these arguments
there is reason to believe that the nearest neighbor surroundings of the
molecules provide the symmetry breaking necessary for the activation of
the forbidden modes.  This mechanism is supported by the fact that
many of the weakly active modes exhibit a sharp change when cooled
through the orientational ordering phase transition.\cite{kamaras,c60modes}
On the other hand the characteristic temperature dependence of the
``defect orientation" fraction \cite{david} should lead to a gradual
decrease in the intensity of the weakly active IR lines originating
from the {\it gerade} modes (in particular Raman modes), and this was
not seen in our experiments.\cite{c60modes}

To summarize, we have shown that isotopic symmetry reduction is {\em not}
the activating mechanism for symmetry forbidden vibrational modes in
C$_{60}$ as has been previously proposed.  We have shown that crystal
field effects can explain the weak modes both above and below the
orientational ordering transition temperature, but further experimental
and theoretical work is required to fully understand the nature of the
symmetry breaking in fullerene crystals.

\acknowledgements

We are grateful to P.B. Allen, S. Pekker, P. Stephens, A. Rosenberg and
G. Sawatzky for enlightening conversations.  This work has been supported
by NSF grants DMR 9202528 (MCM,LM) and DMR 9118414 (JF), and from
CNRS through the Groupe de Recherche 1019 (JG, PB, JML).

\begin{figure}
\caption{Plot of the probability function that a C$_{60}$ molecule has n
$^{13}$C and (60-n) $^{12}$C atoms for the natural abundance of $^{13}$C
and for 8\% enriched samples as measured in the present study.}
\label{dist}
\end{figure}

\begin{figure}
\caption{IR spectra of C$_{60}$ single crystals; the two upper curves have
been vertically offset for clarity.  The top spectra is from a crystal made
from natural abundance $^{13}$C C$_{60}$.  The middle curve is the spectra
of an 8\% $^{13}$C-enriched C$_{60}$ crystal.  The bottom curve is generated
from the top curve using a simple model of frequency shift and the mass
distribution given from Figure 1 for 8\% enrichment.}
\label{spectra}
\end{figure}

\begin{figure}
\narrowtext
\caption{Calculated IR intensities of vibrational modes in C$_{60}$ with
1.108\% and 8\% isotopic substitution.  The four IR-active $F_{1u}$ modes
have intensities of $\sim 2000$ which are well off scale; the scale is chosen
so that the weakly activated symmetry-forbidden modes can be seen.}
\label{theory}
\end{figure}

\end{document}